\documentstyle[12pt,epsf]{ioplppt}

\newcommand{\be}{\begin{equation}}
\newcommand{\ee}{\end{equation}}
\newcommand{\bea}{\begin{eqnarray}}
\newcommand{\eea}{\end{eqnarray}}
\newcommand{\xiimu}{\xi_i^{\mu}}
\newcommand{\vximu}{\vec{\xi}^{\mu}}
\newcommand	{\klm}[1]     {{ \left( {#1} \right) }}

\newcommand	{\skl}[1]     {{ \left\langle{#1} \right\rangle}}
\newcommand	{\btr}[1]     {{ \left|{#1} \right|}}
\begin{document}
\jl{7}
\title{Learning by dilution in a Neural Network}
\author{B L\'opez and W Kinzel}
\address{Institut f\"ur Theoretische Physik, Universit\"at W\"urzburg,
Am Hubland, \\ D-97074~W\"urzburg}
\begin{abstract}
A perceptron with $N$ random weights can store of the order of 
$N$ patterns by removing a fraction of the weights without changing
their strengths. The critical storage capacity as a function of the 
concentration of the remaining bonds for random outputs and for 
outputs given by a teacher perceptron is calculated.
A simple Hebb--like dilution algorithm is presented which in the 
teacher case reaches the optimal generalization ability.
\end{abstract}



\medskip

\section{Introduction}

Neural networks are able to learn from examples and to find
an unknown rule. Storage capacities and generalization abilities
have been calculated for a variety of network architectures within the
framework of statistical mechanics \cite{1}. Special interest has
been devoted to diluted networks, where only a fraction of the neurons
is connected \cite{2,3}. The most popular example of a diluted network
which appears in Nature is the human brain. Every neuron is connected 
to roughly 10000 others, whereas their total number is about $10^{7}$ 
times larger. Theoretical studies indeed showed, that the effective
storage capacity per neuron in diluted systems can be substantially 
larger than in undiluted networks \cite{4}.

So far, dilution of synapses has been considered in addition to
the usual dynamical modification of the bonds, which takes place
in the learning phase \cite{3}.  
Motivated by biological observations
\cite{5}, which indicate that at early stages of development
of the brain, synapses are removed if their
strength is not appropriate, we address the question whether it is possible
to store patterns in a network with randomly chosen coupling strengths
only by removing a fraction of these bonds without changing their
strength. This is a nontrivial task,
since given a specific set of patterns it is a priori not clear which
of the bonds have to be removed. Previous studies \cite{8} have considered
a learning algorithm which removes weights that are frustrated in at least
one of the patterns. However, this simple method removes too many weights,
hence the storage capacity increases with $\log N$ only.
In this paper we show, that it is possible to learn of the order of $N$
patterns perfectly and to generalize by removing a fraction of the weights.
We focus on 
the perceptron, as it is the simplest network for which a tractable
calculation is feasible. The same procedure however should be applicable
to general network classes such as multi-layer perceptrons or attractor
networks.

The paper is organized as follows: In section two we introduce the model
and calculate the critical storage capacity for a random input--output relation
following the standard statistical mechanics approach established by Gardner 
\cite{6}. Section three examines the properties of a perceptron which learns
from an undiluted teacher perceptron. A simple Hebb--like dilution algorithm
that reaches the optimal generalization ability is presented in section four.
In the last section we close with a summary.

\section{The model for random input--output relations}

The diluted perceptron classifies an input pattern $\vximu  $ according
to:
\be
\label{class}
\sigma^{\mu} = \mbox{sgn}\klm{\frac1{\sqrt{N}}\ \sum_{i=1}^N c_i J_i \xiimu} \ .
\ee
Where  $J_i$  are the components of the weight vector drawn at random
from a distribution $P\klm{J_i}$. The $c_i$ are binary decision variables 
which can take the values 0 or 1 and determine whether the $i$th coupling 
$J_i$ has been removed or is kept, respectively.
For a given set of input--output pairs  $\{ \vximu  ,s^{\mu} \}_{\mu = 1}^p$
the classification is correct if:
\be
\label{correct}
s^{\mu} \frac1{\sqrt{N}}\ \sum_{i=1}^N c_i J_i \xiimu \geq 0 \qquad,
\qquad \forall\ \mu = 1,\ldots ,p \ .
\ee
We suppose that the inputs $\xiimu$ are drawn at random from a
distribution with zero mean and unit variance and we choose
$s_{\mu} = \pm 1$ with
equal probability and independently of the inputs. 
The concentration of remaining bonds is defined to be
$c = N^{-1} \sum_{i=1}^N c_i$ and thus lies between 0 and 1. 

We are interested whether the maximum number of patterns $p_{max}$,
which are correctly classified, can be of the order of their input
dimension $N$, resulting in a critical storage capacity of 
$\alpha_c = p_{max}/N$, for 
a fixed value of the concentration $c$ of remaining bonds in the thermodynamic
limit $(N \rightarrow \infty)$. Let us first consider the extreme cases.
For $c=0$ all bonds have been removed and no classification is possible, so
that $\alpha_c(c=0) = 0$. For $c=1$ all bonds are present and the 
classification is at random, so that $\alpha_c(c=1) \rightarrow 0$ as
$N \rightarrow \infty$. For intermediate values of $c$ we shall calculate
$\alpha_c(c)$ using Gardner's phase space approach \cite{6,7}.

Form the technical point of view the problem is related to the
Ising perceptron \cite{7,16} and other discrete models \cite{9}, 
as well as to the knapsack problem \cite{10,11}, where
also binary dynamical variables appear. Note, that in contrast to the
common approach where the couplings $J_i$ are the dynamical variables,
here they represent in addition to the patterns a quenched disorder
which has to be averaged out.

For a fixed concentration $c$, the number of allowed configurations
according to \eref{correct} is given by
\be
\label{number}
{\cal N}\klm{c} = \sum_{\{c_i\} }\ \prod_{\mu = 1}^p 
              \Theta \klm{
              s^{\mu} \frac1{\sqrt{N}}\ \sum_{i=1}^N c_i J_i \xiimu - \kappa
                         }\ 
              \delta_{\mbox{\tiny  Kr}} \klm{
                     \sum_{i=1}^N c_i  -  c N 
                      } \ .       
\ee
We introduced, as usual, the stability parameter $\kappa$ which should be
positive.
The corresponding entropy per bond of the microcanonical ensemble follows 
from
\be
S\klm{c} = \frac1N\ \skl{\skl{ \mbox{ln}\ {\cal N}\klm{c} }} = 
       \lim_{n \rightarrow 0}\ \frac1{n N} 
       \mbox{ln} \skl{\skl{ {\cal N}^n\klm{c} }}
       \ ,
\ee
where the last equality results from the replica trick. The quenched
averages $\skl{\skl{\ldots}}$ have to be performed over the distributions
of the patterns, outputs and the couplings. Following the steps of the 
calculation by Gardner and Derrida (1988) one can rewrite the replicated number of
configurations using the integral representation of the theta function
in \eref{number}.
The averages over the pattern and output distributions lead to the exponential 
factor
\be
\label{expon}
\skl{\prod_i \exp \klm{ - \frac{J_i^2}{2 N} \sum_{\mu} 
                         \klm{\sum_{\alpha} c_i^{\alpha} x_{\mu}^{\alpha}
                         }^2
                       }
     }_{ \{ J \} } \ .                 
\ee
Here $x_{\mu}^{\alpha}$ are the conjugate variables to the local fields
$\lambda_{\mu}^{\alpha}$ and $\alpha$ denotes the replica index running
from 1 to $n$. The average over the couplings has still to be done.
A straightforward evaluation however, leads to an expression which cannot
be rewritten in terms of an exponential, as it is convenient, for 
the argument of the exponent in \eref{expon} is  not necessarily
infinitesimal due to the sum over all patterns. Instead, we introduce
at this point the order parameters
\bea
 \label{qalpha}
q^{\alpha \beta} &=& \frac1N\ \sum_{i=1}^N J_i^2 c_i^{\alpha} c_i^{\beta} 
 \qquad ( \alpha ,\beta = 1,\ldots , n \ ; \  \alpha < \beta) \\
 \label{Qalpha}
Q^{\alpha} &=& q^{\alpha \alpha} = \frac1N\ \sum_{i=1}^N J_i^2 c_i^{\alpha} 
  \qquad (\alpha = 1,\ldots , n)
\eea
and leave the average over the distribution of the couplings for the
integral representations of the delta functions fixing $q^{\alpha \beta}$
and $Q^{\alpha}$. In addition one has a third order parameter $E^{\alpha}$
which fixes the concentration $c$.
We seek for a replica symmetric solution,
i.e. $q^{\alpha \beta} = q$ , $Q^{\alpha} = Q$  and $E^{\alpha} = E$.
Carrying out the sum over the $c_i^{\alpha}$ and using the saddle--point method  
we obtain for the entropy in the thermodynamic limit: 
\bea
\label{entropy}
\fl  S\klm{c} =   \alpha \int Dt\ \ln \mbox{H} \klm{
             \frac{\kappa + \sqrt{q} t}{\sqrt{Q - q}}  }
                    + \frac12 F q - \frac12 f Q + \frac12 c E  
                    \nonumber \\
 +        \int dJ P\klm{J} \int Dt\
          \ln \klm{ 1 + \exp \klm{ \sqrt{F} \btr{J} t + 
                                    \frac12 \klm{f J^2 - F J^2 - E}
                                            }
           }  \ ,
\eea                     
where $F$ and $f$ are the conjugate order parameters to $q$ and $Q$
respectively and we have used the notations:
\be 
Dt \equiv dt\ \frac{\exp \klm{-\frac12 t^2} }{\sqrt{2 \pi}}
\qquad , \qquad
\mbox{H}(x) = \int_x^{\infty} Dt \ .
\ee

\begin{figure}
\hbox{
\epsfysize 12cm
\hbox to \hsize{\hss
\epsfbox{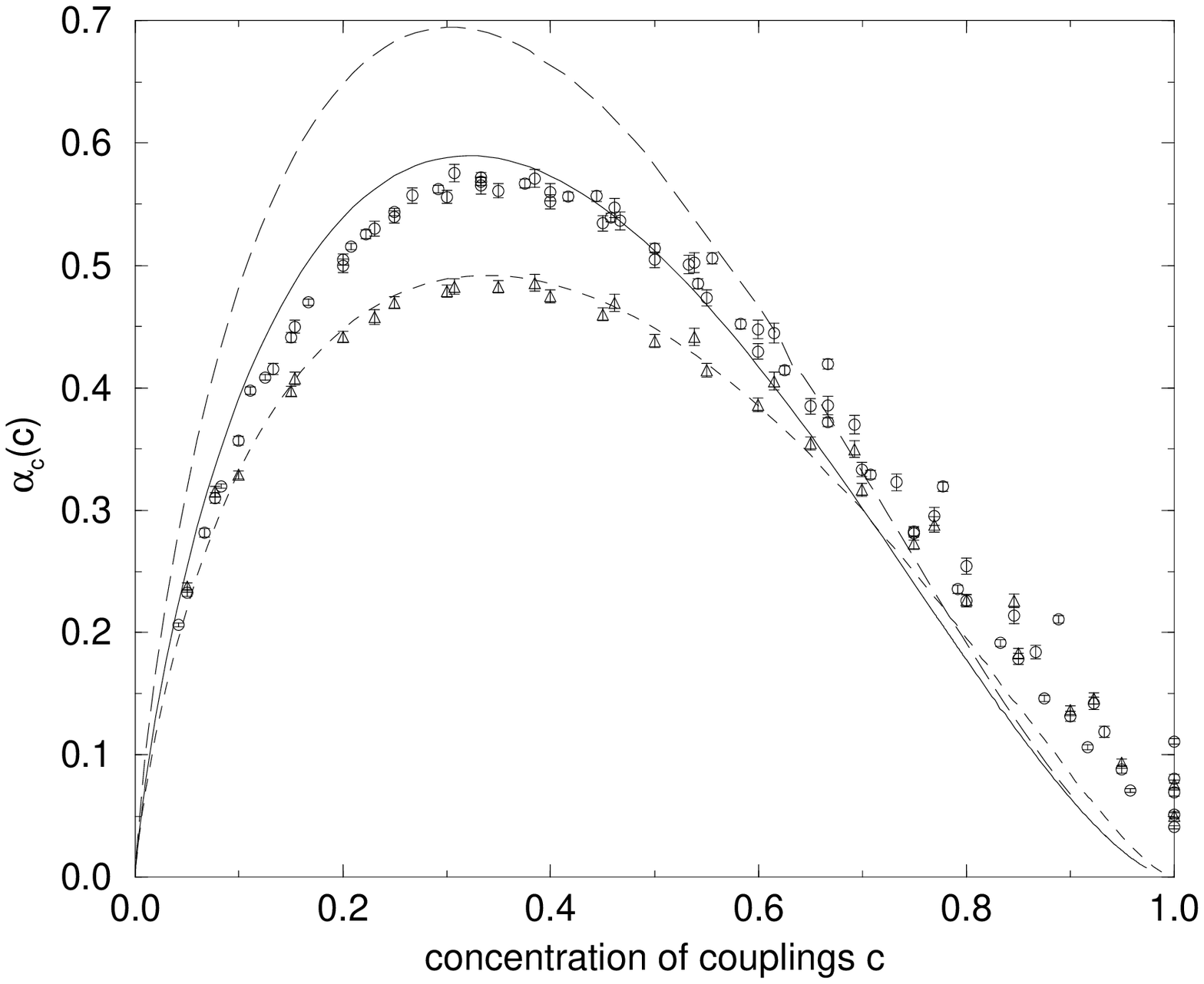} \hss}}
\caption{The critical storage capacity $\alpha_c$ as a function of the
         concentration $c$ for $\kappa = 0$ and random outputs. The solid
         line represents the zero entropy solution for $\btr{J_i}=1$. 
         The dots with error bars are results from complete enumerations
         of systems with sizes $9 \leq N \leq 24$. The long dashed
         curve is the corresponding AT--line beyond which the RS--solution becomes
         locally unstable. The dashed curve is the critical storage 
         capacity for $J_i$ drawn from a normal distribution.}
\label{jrand}
\end{figure}

For the distribution of the couplings $P\klm{J_i}$ we will focus on 
two cases: $\btr{J_i}=1$ (note, that the entropy \eref{entropy} depends
only on the absolute value of $J_i$) and $J_i$ drawn from a normal
distribution i.e. $dJ P\klm{J} = DJ$. In the first case it follows that
$Q=c$ from \eref{Qalpha} and the definition of $c$. We solve the saddle
point equations $\partial S / \partial q = \partial S / \partial F
=  \partial S / \partial E = 0$. The critical storage capacity as determined
by Gardner and Derrida \cite{7} would be reached as $q$ approaches $c$.
This however, leads to a negative entropy of the system, as frequently 
observed in discrete problems \cite{13}.
We therefore identify the critical storage capacity as the value of 
$\alpha$ at which the entropy \eref{entropy} vanishes, as it has become 
the standard way by now \cite{9,11,14}. 
The resulting curve for the critical storage capacity $\alpha_c$ is 
shown in \fref{jrand} as a function of the concentration $c$. As expected
$\alpha_c$ vanishes at both extremes of $c$. We observe a maximum  of 
$\alpha_c \approx 0.59$ at $c \approx 0.32$. This means, that about 
2/3 of the bonds have to be removed in order to reach the maximal value
of $\alpha_c$. It is somewhat surprising, since as a function of $c$, the
maximal number of configurations $\{c_i\}$ lies at $c=0.5$ and is exponentially
larger in $N$ than for any other value of the concentration. In 
section~\ref{algorithm} we will come back to this point.

It is worth to note, that the case $\btr{J_i}=1$ can be mapped onto the
Ising perceptron with couplings 0 or 1, as in \eref{correct} one can define
the new patterns $\chi_i^{\mu} = J_i \xiimu$ and view the $c_i$ as the 
couplings. The distribution of the $\chi_i^{\mu}$ has also zero mean and
unit variance. The critical storage capacity for the (0,1) Ising perceptron 
has been calculated by Gutfreund and Stein \cite{9} with the zero-entropy (ZE) 
ansatz and was found to be 0.59 in agreement with the value found here 
at the maximum.

We performed an analysis of the local stability of the replica symmetric
(RS) solution according to de Almeida and Thouless \cite{15}
and obtained the curve
denoted as AT--line in \fref{jrand}. For values of $\alpha$ that lie
above this curve the RS solution is locally unstable. 
Our result for $\alpha_c(c)$ lies below the AT--line for all $c$ and is 
therefore locally stable. Nonetheless, global stability is not assured.
For this reason we also performed complete enumerations of all possible
dilution vectors for finite systems. The dots with their corresponding 
error bars are results for systems with $9 \leq N \leq 24$. For $c=1$
finite size effects lead to $\alpha_c \sim N^{-1}$, since the probability of
classifying one pattern correctly by chance is 1/2. In contrast, for
values of $c$ around the maximum, the numerical results seem to underestimate
the theoretical values. A finite size scaling analysis for $c=1/3$ gives
the extrapolated value of $\alpha_c(1/3) = 0.586 \pm 0.004$ for 
$N \rightarrow \infty$ in agreement with the RS--solution (0.58935)
at the same concentration. The general shape of the curve is well
confirmed by the numerical results.

In the case where the $J_i$ are drawn at random from a normal distribution
the picture changes quantitatively. Now, for a fixed $i$, 
$\skl{J_i \xiimu} = 0$ as before, but
$\skl{\klm{J_i \xiimu}^2} = J_i^2 \skl{{\xiimu}^2} = J_i^2$ which is in general 
different from unity as in the previous case. As a consequence $Q$ is
different from $c$ and we have to solve the additional saddle point equations
$\partial S / \partial Q = \partial S / \partial f = 0$. The ZE condition
yields for $\kappa = 0$ the critical storage capacity depicted also in 
\fref{jrand}. Similarly, the curve has a maximum at $c \simeq 0.34$, the
critical capacity however is lowered over a wide range of the concentration
with respect to the binary case. Our interpretation for this effect is that
in the Gaussian case 
the dilution variables are mainly used to remove the large couplings
$(\btr{J_i} > 1)$ and only few of them remain for learning. Therefore the
storage capacity $\alpha_c$ is lower than for the binary weights.
The order parameter $Q$ measures the
effective size of the remaining components $J_i$. For all $c$ the RS--solution
gives $Q < c$, supporting the above argument. In addition we measured 
the probability
distribution of the size of remaining couplings in complete enumerations,
and found, that large couplings are likely to be removed. The values of
$\alpha_c$ for finite $N$ are displayed in \fref{jrand} as well.

\section{Learning with a teacher}
\label{teacher}

A teacher perceptron $\vec{B}$ classifies a pattern $\vximu  $ according
to:
\be
\label{clteach}
s^{\mu} = \mbox{sgn}\klm{\frac1{\sqrt{N}}\ \vec{B} \cdot \vximu  }
\ee
We choose a teacher vector which is not diluted and which has the 
normalization $\vec{B}^2 = N$. A transition to perfect generalization through
dilution cannot be expected like in other discrete systems where the structure
of teacher and student coincides \cite{23}. The student 
perceptron with components $c_i J_i$ can only remove part of 
its weights in order to learn perfectly a set of examples given by 
\eref{clteach}, resulting in a finite storage capacity. 

A straightforward evaluation of the entropy under the RS assumption yields:
\bea
\label{entroteach}
\fl  S\klm{c} = 2 \alpha \int Dt\ 
             \mbox{H} \klm{\frac{R\ \sqrt{Q}}{\sqrt{q - Q R^2}}\ t}
             \ln \mbox{H} \klm{
             \frac{\kappa + \sqrt{q} t}{\sqrt{Q - q}}  }
              + \frac12 F q\ - \frac12 f Q\ + \frac12 c E\ + \frac12 G R\ + 
                    \nonumber \\
\fl       \int dB P_{\mbox{\tiny T}}\klm{B} \int dJ P\klm{J} \int Dt
          \ln \klm{ 1 + \exp \klm{ \sqrt{F} \btr{J} t + 
                                    \frac12 \klm{f J^2 - F J^2 - E - G B J}
                                            }
           }  \ ,
\eea                     
where the overlap $R = \sum_i c_i J_i B_i / \klm{\sqrt{Q} N}$ between 
diluted student and teacher, and its conjugate $G$ has been introduced.
$P_{\mbox{\tiny T}}\klm{B}$ is the distribution of the teacher components. 

\begin{figure}
\hbox{
\epsfysize 8cm
\hbox to \hsize{\hss
\epsfbox{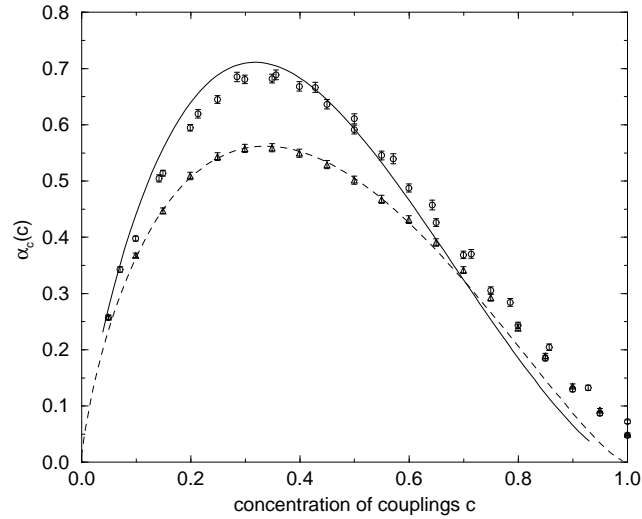} \hss}}
\caption{The critical storage capacity $\alpha_c$ as a function of the 
         concentration $c$ for $\kappa = 0$ and outputs from an undiluted
         teacher perceptron. The upper curve is for $\btr{J_i}=\btr{B_i}=1$
         and the lower for $J_i$ and $B_i$ both Gaussian. The dots are 
         numerical results from complete enumerations of systems with 
         sizes up to $N=20$.}
\label{tuncorr}
\end{figure}

We focus again on the two cases where $\btr{J_i} = \btr{B_i} = 1$ or both
chosen independently at random from a normal distribution. The corresponding
critical storage capacity $\alpha_c$ determined  with the ZE condition
is shown in \fref{tuncorr} as a function of the concentration of remaining
bonds. Once more, we observe a maximum  of $\alpha_c$ at $c \approx 1/3$.
The critical storage capacity is higher than for random outputs indicating
that the problem, although unlearnable, is easier with examples from a teacher.

An important point to note is that the generalization ability defined as the
probability to classify a new unseen pattern correctly (as the teacher)
is poor compared to the value which could be achieved by {\it intelligent}
dilution. For $c=0.3$, $R \simeq 0.32$ in the binary case $\btr{J_i} = 
\btr{B_i} = 1$, whereas an overlap of $R = \sqrt{0.3} \simeq 0.55$ could
be possible according to the following argument. The product $B_i J_i$
should be $+1$ for as many sites as possible in order to maximize $R$.
Since $B_i$ and $J_i$ are drawn at random they will coincide in $N/2$ of
the cases for $N \rightarrow \infty$. For $c \leq 0.5$ we choose all those
$c_i = 1$ for which $B_i J_i = +1$ up to a total number of $c N$, so that
$R_{max} = c N / \klm{ \sqrt{c} N} = \sqrt{c}$. If c is larger than $0.5$
then we also have to add up $\klm{ c N - N/2}$ times a value of $-1$ and
$R_{max} = \klm{N/2 - c N +N/2}/\klm{ \sqrt{c} N} = \klm{1-c}/\sqrt{c}$.
Perfect storage without errors lowers the overlap $R$, an effect known as
{\it overfitting}, which can be overcome by allowing a finite training error
(see section~\ref{algorithm}).  

Up to now, we have assumed, that student $\vec{J}$ and teacher 
$\vec{B}$ are uncorrelated before the dilution. In biological systems
however, we would rather expect to find synaptical structures, which
are already {\it prepared} for a specific task before the learning process
starts. In our model we can mimic it by allowing an initial positive overlap
$R_0$, and therefore certain similarity, between teacher and student.
Let us choose
\be
P\klm{J} = \frac{1+R_0}2\ \delta\klm{J-B}\ +\ \frac{1-R_0}2\ \delta\klm{J+B}
\ee
with $0 \leq R_0 \leq 1$. In $(1+R_0) N /2$ of the cases the components
of teacher and student will coincide and in $(1-R_0) N /2$ they will be 
opposed. Since we are not allowed to change the values of $J_i$, but at best
to remove the bond, we will not reach perfect generalization even for
large $R_0$. The same would even hold if we chose a diluted teacher.

\begin{figure}
\hbox{
\epsfysize 8cm
\hbox to \hsize{\hss
\epsfbox{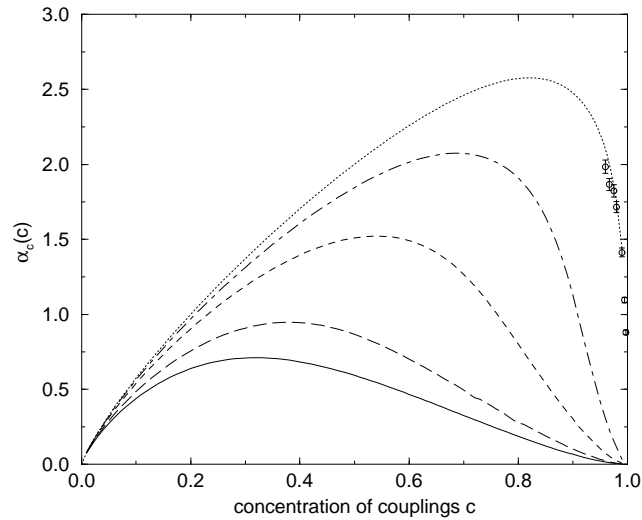} \hss}}
\caption{The critical storage capacity $\alpha_c$ as a function of the
         concentration $c$ for $\kappa = 0$ and outputs from a teacher 
         with $\btr{B_i}=1$ for different values of the initial overlap
         $R_0$ between $\vec{J}$ and $\vec{B}$. $R_0 = 1.0, 0.9, 0.7,
         0.3, 0.0$ (from top to bottom). The dots are numerical results 
         from complete enumerations of systems with sizes $25 \leq N 
         \leq 400$ and $c$ close to 1.}
\label{tcorr}
\end{figure}

We have calculated the critical storage capacity for $\btr{B_i}=1$ as a 
function of $R_0$ and $c$ and find the results plotted in \fref{tcorr}.
For $R_0=0$ we recover the uncorrelated case, whereas with increasing $R_0$
the storage capacity is enhanced for all $c$. At the same time the maximum
of the curve moves towards higher values of the concentration, as less bonds
have to be removed in order to mimic the teacher. For $R_0=1$ teacher and student 
are identical before dilution. Although this situation is of less relevance
from the biological point of view, it offers an interesting physical
solution. If a fraction of the bonds is now removed, we obtain a finite storage
capacity. For increasing concentration $c$, we would expect the capacity
to increase at the same time. Above $c \simeq 0.82$ however, we find, that
it decreases and finally tends to zero for $c \rightarrow 1$. At $c=1$
we have $\alpha_c = \infty$ per definition, hence $c=1$ is a singular point. 
This surprising behaviour may
be understood if we look at the annealed approximation \cite{22} for the
entropy
\be
\label{sann}
S_{ann}\klm{c} = \frac1N\ \mbox{ln} \skl{\skl{ {\cal N}\klm{c} }}.
\ee
The averaged number $\skl{\skl{ {\cal N}\klm{c} }}$ of allowed configurations 
for fixed $c$ is simply given by the total number ${N \choose c N}$
times the probability that $p$ patterns are classified correctly:
\be
\label{naver}
\skl{\skl{ {\cal N}\klm{c} }} =  {N \choose c N} 
                                 \klm{1-\frac1{\pi} \arccos R}^p\ .
\ee
For $R_0 =1$ we have $R=\sqrt{c}$ and using the ZE--condition one obtains
in the thermodynamic limit for the critical storage capacity in the annealed
approximation
\be
\alpha_{ann}\klm{c} = \frac{c\ \mbox{ln} c + \klm{1-c}\ \mbox{ln} \klm{1-c}}
                           {\mbox{ln} \klm{1 - \frac1{\pi} \arccos \sqrt{c}}}\ ,
\ee                                                            
which for $c \rightarrow 1$ results in
\be
\alpha_{ann}\klm{c \rightarrow 1} \rightarrow \lim_{c \rightarrow 1} - 4 \pi
        \sqrt{1-c}\ \mbox{ln} \sqrt{1-c} \rightarrow 0.
\ee
Since $\alpha_{ann}\klm{c}$ is an upper bound for $\alpha_c\klm{c}$, the 
critical storage capacity has to decrease to zero as well when $c$ tends to 1.
From \eref{naver} we see, that although the probability of classifying one
pattern correctly tends to one for $c \rightarrow 1$, at the same time the 
total number of dilution vectors decreases rapidly, such that the averaged
number of allowed configurations is no longer exponentially large in $N$.
Perfect storage of patterns is different from optimal generalization, which
in this case becomes better the closer $c$ is to 1.
In \fref{tcorr} we also included results from complete enumerations of 
systems with $25 \leq N \leq 400$ for $R_0 =1$ and $c$ close to 1. They confirm
that $\alpha_c$ decreases in this region.

\section{A simple Hebb--like dilution algorithm}
\label{algorithm}

As we have seen in the previous sections, Gardner's method is very powerful
when asking if there exists, on average, a set of $c_i$ such that all perceptron 
conditions \eref{correct} are satisfied, but it does not provide us with 
the corresponding dilution vector for a specific set of patterns $\vximu  $, outputs 
$s^{\mu}$ and couplings $J_i$.
The development of a learning , or in our case dilution algorithm, 
is an independent task, which in the case of binary variables 
$c_i=0,1$ becomes extremely difficult and compares with the binary 
perceptron problem or the knapsack problem. In the worst case, the 
number of computational steps towards the optimal solution scales 
exponentially with the size $N$ of the system. The most successful 
approaches try to find the global minimum of a 
properly defined energy function, which penalizes the violation 
of constraints, by using sequential descent \cite{17} 
or simulated annealing \cite{18} strategies. 
Although an algorithm based on mean field annealing has proven 
to be very effective in finding solutions to the knapsack problem 
\cite{10}, none of the known techniques yields the critical 
values for the storage capacity predicted by Gardner calculations. 
Typically, in large systems, the solutions still violate a finite 
fraction of the imposed constraints. 
 
In view of these general difficulties, we cannot expect to find the 
optimal dilution vector $\vec{c}$, which allows us to store perfectly 
$\alpha_c N$ many patterns for large $N$, within reasonable time. 
We rather present here a simple dilution algorithm, 
which gives us an insight into the basic properties of the solutions 
and has optimal generalization ability for $\alpha \rightarrow \infty$. 
 
Our aim is to fulfill all constraints \eref{correct} by removing 
a fraction $\klm{1-c} N$ of the bonds $J_i$. If we think of the terms 
$J_i \xiimu  s^{\mu}$ as matrix elements $a_{i \mu}$ of a $N \times p$ 
matrix, then we want all $p$ vertical sums $\sum_{i=1}^N a_{i \mu}\ 
(\mu = 1, \ldots, p)$ to be positive. For this purpose, we are allowed 
to remove $\klm{1-c} N$ rows of the matrix. The idea is to remove those,
 which contain many negative elements $a_{i \mu}$, since these 
contribute in many vertical sums negatively. Let us take away all 
rows $i$ with horizontal sums $\sum_{\mu =1}^p a_{i \mu}$ smaller than 
a threshold $h$, so that 
\be 
\label{dilalg} 
c_i = \theta \klm{ \frac1{\sqrt{N}} \sum_{\mu=1}^p J_i \xiimu  s^{\mu} - h} 
\ee 
The larger $h$, the more $c_i$ will be zero, leading to a lower 
concentration $c$. From a different perspective one can view \eref{dilalg} 
as comparing the Hebb couplings 
$H_i = \sum_{\mu =1}^p \xiimu  s^{\mu} /\sqrt{N}$ 
of the problem (Hopfield 1982) with $J_i$, the ones imposed at random. 
If their product $H_i J_i$ is larger than the threshold $h$, then $J_i$ 
is accepted as coupling strength. As $h$ becomes more and more positive, 
this is only the case if $H_i$ and $J_i$ agree in their sign. 
 
\begin{figure}
\hbox{
\epsfysize 8cm
\hbox to \hsize{\hss
\epsfbox{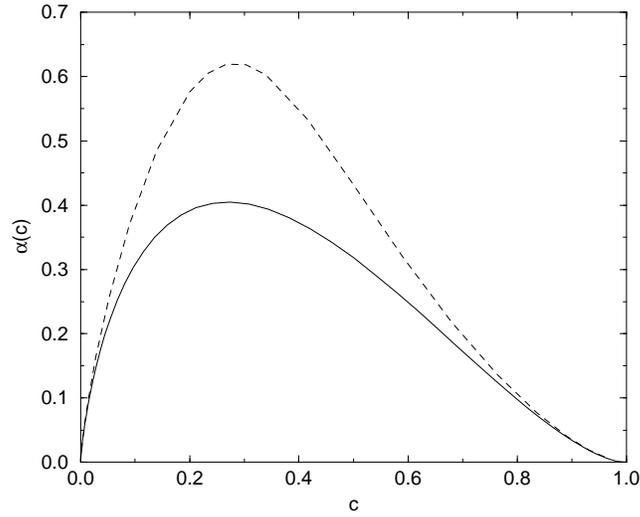} \hss}}
\caption{The storage capacity $\alpha$ as a function of the concentration
         $c$ for $\btr{J_i}=1$, $\kappa = 0$ and $A=1\ 
         (\epsilon_{\mbox{\tiny L}} \simeq 15.9 \%)$. The lower curve is
         for randomly chosen outputs and the upper curve for outputs from
         a teacher with $\btr{B_i}=1$.}
\label{jbinrnalg}
\end{figure}

Let us now give a simple derivation of the critical storage capacity 
for random input--output pairs and $\btr{J_i} = 1$, which results from 
\eref{dilalg} by allowing a certain percentage of errors. For simplicity, 
suppose that $\xiimu  = \pm 1$ with equal probability. Then, 
$a_{i \mu} = \pm 1$ with equal probability and the horizontal and vertical 
sums have, in the limit $N \rightarrow \infty$, a Gaussian distribution 
with zero mean and variance $p$ and $N$, 
respectively. According to \eref{dilalg} we remove all horizontal sums 
which are smaller than $h$. The resulting concentration is 
$c = \int_{h/\sqrt{\alpha}}^{\infty} Dz = \mbox{H} \klm{h/\sqrt{\alpha}}$ 
and the new mean of the horizontal sums is 
$\skl{\mbox{HS}} = \sqrt{p} \exp \klm{-\case12 h^2/\alpha}/\sqrt{2 \pi c^2}$. 
The new vertical sums have still a Gaussian distribution, but now with 
mean $\skl{\mbox{VS}} = c \skl{\mbox{HS}}/\alpha  $ and variance 
$c N$ for $N \rightarrow \infty$. The fraction of errors is thus 
equal to the integral over the Gaussian tail below zero: 
\be
\label{errrand} 
\fl
\mbox{Learning Error} \equiv \epsilon_{\mbox{\tiny L}} = 
                      \int_{- \infty}^0 \frac1{\sqrt{2 \pi}} 
                      \frac1{\sqrt{c N}} 
                      \exp \klm{-\frac12 \frac{\klm{z-\skl{\mbox{VS}}}^2} 
                                             {c N}  } 
                       = \mbox{H} \klm{\frac{\skl{\mbox{VS}}}{\sqrt{c N}}} 
\ee 
For a fixed error $\epsilon_{\mbox{\tiny L}} = \mbox{H} \klm{A}$ and fixed 
concentration $c$, we obtain for the storage capacity: 
\be 
\label{alfalg} 
\alpha \klm{c,A} = \frac{ \exp \klm{ - \klm{\mbox{H}^{-1}\klm{c}}^2 }} 
                        { 2 \pi c A^2 } 
\ee 
with $\mbox{H}^{-1}\klm{x}$ the inverse function of $\mbox{H}\klm{x}$. 
\Fref{jbinrnalg} shows the resulting $\alpha$ for $A=1\ 
(\epsilon_{\mbox{\tiny L}} \simeq 15.9 \%)$ as a function of $c$. The maximum  
of $\alpha$ is reached at $c \approx 0.27$, which is not too far from 
$0.32$, the concentration at which the maximal $\alpha_c$ was obtained 
according to the Gardner calculation with the ZE condition 
(see \fref{jrand}). Also the shape 
of the curve is similar, for different values of $\epsilon_{\mbox{\tiny L}}$ 
(or $A$), $\alpha\klm{c,A}$ is simply rescaled. 
At first sight one would expect that as $h$ increases, also the mean of
the vertical sums $\skl{\mbox{VS}}$ increases, leading to a lower learning
error. This however is prevented by two effects. First, as $\skl{\mbox{HS}}$
increases, $\skl{\mbox{VS}}$ does not so necessarily, since 
$\skl{\mbox{VS}} \sim c \skl{\mbox{HS}}$ and $c$ is lowered dramatically
with increasing $h$. As a result, the maximal storage capacity would be at
$c = 0.5$ for fixed $\epsilon_{\mbox{\tiny L}}$. The second effect is, that
the width of the distribution of vertical sums is proportional to $\sqrt{c}$,
lowering the learning error for smaller values of $c$. If $c$ is too small
however, the gain is compensated by the exponential factor in $\skl{\mbox{HS}}$,
which tends to zero. As a consequence, the maximum storage capacity for fixed
$\epsilon_{\mbox{\tiny L}}$ lies at a value of $c$ somewhat smaller than $0.5$. 

\begin{figure}
\hbox{
\epsfysize 8cm
\hbox to \hsize{\hss
\epsfbox{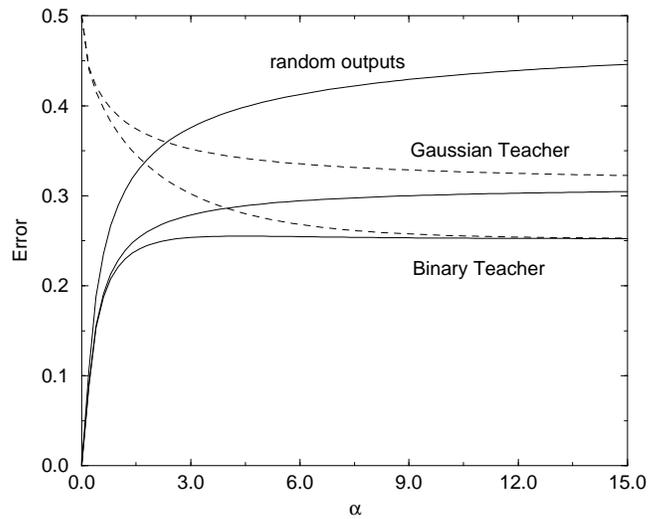} \hss}}
\caption{Learning error $\epsilon_{\mbox{\tiny L}}$ (solid lines) and 
         generalization error $\epsilon_{\mbox{\tiny G}}$ (dashed lines)
         as a function of $\alpha$ for $h=0\ \klm{\Leftrightarrow c=0.5}$ 
         and $\btr{J_i}=1$ for random outputs, a binary teacher 
         and a Gaussian teacher.}
\label{epsalg}
\end{figure}

In \fref{epsalg} the learning error $\epsilon_{\mbox{\tiny L}} \klm{c,\alpha}$
is plotted as a function of $\alpha$ for $h=0\ \klm{\Leftrightarrow c=0.5}$.
For small $\alpha$, $\epsilon_{\mbox{\tiny L}}$ is small, as typical for the
Hebb couplings and tends to 0.5 for $\alpha \rightarrow \infty$. 

A more interesting quantity is the generalization error
$\epsilon_{\mbox{\tiny G}}$, defined as the probability to classify correctly
a new pattern ${\vec{\xi}}^{\ 0}$, which does not belong to the training set. For a
random input--output relation $\epsilon_{\mbox{\tiny G}} = 0.5$, since the
classification $s^0$ of ${\vec{\xi}}^{\ 0}$ is at random. In the presence of a teacher
$\vec{B}$ however, we can expect to reach a lower generalization error. A
straightforward evaluation (see e.g. \cite{20}) yields for
$\btr{J_i} = 1$:
\bea
\epsilon_{\mbox{\tiny L}} \klm{c,\alpha} =
      2 \int_0^{\infty} Ds\ \mbox{H} \klm{\frac{s R \sqrt{c} +q}
                                              {\sqrt{c}\ \sqrt{1-R^2}}} \\
\epsilon_{\mbox{\tiny G}} \klm{c,\alpha} = \frac1{\pi} \arccos R \,
\eea
with
\bea
c = \frac12\ \skl{\mbox{H} \klm{ h_- }
               + \mbox{H} \klm{ h_+ }}_{B_i}  \\
R = \frac1{2 \sqrt{c}}\ \skl{ B_i\ \klm{
                 \mbox{H} \klm{ h_- }
               - \mbox{H} \klm{ h_+ } } }_{B_i}  
               \label{R}   \\
q = \frac1{2 \sqrt{2 \pi \alpha}}\ \skl{
                 \exp \klm{-\frac12 h_-^2 } +
                 \exp \klm{-\frac12 h_+^2 }  }_{B_i} \,
\eea
where
\be
h_+ =   \frac{h}{\sqrt{\alpha}} + B_i\ \sqrt{ \frac{2\alpha}{\pi} }  
           \qquad , \qquad
h_- =   \frac{h}{\sqrt{\alpha}} - B_i\ \sqrt{ \frac{2\alpha}{\pi} }  \ .
\ee                                   
The average is to be performed over the distribution $P_{\mbox{\tiny T}}\klm{B_i}$
of the teacher components $B_i$. As before, $c$ is the concentration of 
remaining bonds and $R$ the normalized overlap between diluted student and
teacher. The parameter $q$ (not to be mismatched with $q$ defined by
\eref{qalpha}) does not seem to have a direct physical meaning, but in a
certain way it does take into account the randomness which is still inherent
for $R$ less than 1. In the extreme case where we ignore the teacher by
setting all $B_i =0$, we obtain $R=0\ ,\ q = \exp \klm{-\case12 h^2/\alpha}/
\sqrt{2 \pi \alpha}$ and recover for $\epsilon_{\mbox{\tiny L}}$ the expression
for random outputs \eref{errrand}. The above result for 
$\epsilon_{\mbox{\tiny L}}$ and $\epsilon_{\mbox{\tiny G}}$ is similar to
the one obtained with the Clipped--Hebb algorithm for the perceptron
\cite{21}. This does not surprise, as our prescription
\eref{dilalg} is also in some sense a way of {\it clipping} the bonds.

For all $\alpha$ and $c$ we find 
$\epsilon_{\mbox{\tiny L}} < \epsilon_{\mbox{\tiny G}}$
and for $\alpha \rightarrow \infty$ and $c$ fixed, 
$\epsilon_{\mbox{\tiny L}} \rightarrow \epsilon_{\mbox{\tiny G}}$, as it
should.
The most important feature however is, that in this same limit the optimal
generalization error is reached. For $\alpha \rightarrow \infty$ and $c$ 
fixed by choosing $h$ appropriately, we obtain from \eref{R} for 
$\btr{B_i} =1$:
\be
R = \cases{\sqrt{c} &for $c \leq \frac12$  \\
            \frac{1-c}{\sqrt{c}} &for $c > \frac12$ \\ }
\ee
which is exactly the maximal overlap that can be achieved by removing
$\klm{1-c} N$ bonds (see section~\ref{teacher}). For a teacher with
Gaussian $B_i$ we find in the same limit:
\be
R = \frac1{\sqrt{2 \pi c}}\ \exp \klm{- \frac12\ \klm{\mbox{H}^{-1}\klm{c}}^2 }
\ee
which is also the optimal overlap for this case, as can be shown easily.

\section{Conclusion}

We have shown, that it is possible to store information in a neural network
solely by dilution of synapses. Using Gardner's
phase space approach the critical storage capacity of a
perceptron with random coupling vector was calculated 
as a function of the concentration of remaining
bonds for random input--output relations. We found a maximum 
$\alpha_c \simeq 0.6$ of the 
capacity at $c \simeq 1/3$, i.e.\ after 2/3 of the bonds have been removed.
Similar results are obtained if the desired outputs are generated by
an undiluted teacher perceptron, whose coupling vector is uncorrelated
to the initial student vector. In this case perfect learning is possible
up to a critical capacity $\alpha_c(c)$, only. If the initial network has
some preknowledge, i.e.\ if there is a nonzero overlap between teacher
and initial student vector we find, that the maximum of the capacity 
moves towards higher concentrations $c$. 

The problem of finding the subset of couplings which have to be removed
is extremely difficult and compares to problems which belong to the
NP--complete class. Nevertheless, properties of a Hebb--like learning
algorithm, which allows for a finite fraction of errors in the training
set, were calculated. The algorithm reaches for $\alpha \rightarrow \infty$
the maximal overlap between the diluted random student and the undiluted
teacher and thus the lowest possible generalization error. As the Hebb--rule,
it is a local algorithm that accumulates information about the training set
and decides at the end of this batch process which of the couplings are
removed. More desirable would be to find a prescription that {\it on--line}
removes disturbing couplings. In contrast to common on--line learning
algorithms \cite{12}, where infinitesimal changes of the coupling vector
are performed in every time step, here we would remove single couplings.
Whether this procedure can give satisfactory results, similar to those
obtained in the batch process, also for unlearnable rules, remains open
and should be studied in the future.

\ack
We would like to thank M Opper and A Mietzner for useful discussions.
This work has been supported by the Deutsche Forschungsgemeinschaft.

\Figures

\end{document}